\documentclass[conference]{IEEEtran}
\IEEEoverridecommandlockouts
\usepackage{cite}
\usepackage{amsmath,amssymb,amsfonts}
\usepackage{algorithmic}
\usepackage{graphicx}
\usepackage{textcomp}
\usepackage{pdfpages}
\usepackage{xcolor}
\usepackage{booktabs}
\def\BibTeX{{\rm B\kern-.05em{\sc i\kern-.025em b}\kern-.08em
    T\kern-.1667em\lower.7ex\hbox{E}\kern-.125emX}}

\usepackage{xspace}
\usepackage{hyperref}
\usepackage[show]{chato-notes}

\begin{document}

%%%%%%%%%%%%%%%

\newcommand\acronym{ScalOTA\xspace}
\newcommand\station{EVU station\xspace}

\newcommand{\sig}[1]{\langle #1 \rangle}
\newcommand{\msf}[1]{\mathsf{#1}}

\title{\acronym: Scalable Secure Over-the-Air Software Updates for Vehicles}

%\author{\IEEEauthorblockN{DSN Submission \#N}}

\author{\IEEEauthorblockN{1\textsuperscript{st} Ali Shoker}
\IEEEauthorblockA{\textit{RC3, KAUST} \\
%\textit{Computer, Electrical and Mathematical Sciences and Engineering Division (CEMSE),} \\
%\textit{King Abdullah University of Science and Technology (KAUST)}\\
%Thuwal 23955-6900, Kingdom of Saudi Arabia \\
ali.shoker@kaust.edu.sa}
\and
\IEEEauthorblockN{2\textsuperscript{nd} Fernando Alves}
\IEEEauthorblockA{\textit{VORTEX CoLAB} \\
fernando.alves@vortex-colab.com}
\and
\IEEEauthorblockN{3\textsuperscript{rd} Paulo Esteves-Verissimo}
\IEEEauthorblockA{\textit{RC3, KAUST} \\
 paulo.verissimo@kaust.edu.sa}
}

\maketitle
\pagestyle{plain} 

%%%%%%%%%%%%%%%%%%%

\begin{abstract}
Over-the-Air (OTA) software updates are becoming essential for electric/electronic vehicle architectures in order to reduce recalls amid the increasing software bugs and vulnerabilities. Current OTA update architectures rely heavily on direct cellular repository-to-vehicle links, which makes the repository a communication bottleneck, and increases the cellular bandwidth utilization cost as well as the software download latency. In this paper, we introduce ScalOTA, an end-to-end scalable OTA software update architecture and secure protocol for modern vehicles. For the first time, we propose using a network of update stations, as part of Electric Vehicle charging stations, to boost the download speed through these stations, and reduce the cellular bandwidth overhead significantly. Our formalized OTA update protocol ensures proven end-to-end chain-of-trust including all stakeholders: manufacturer, suppliers, update stations, and all layers of in-vehicle Electric Control Units (ECUs). The empirical evaluation shows that ScalOTA reduces the bandwidth utilization and download latency up to an order of magnitude compared with current OTA update systems.

\end{abstract}

\section{Introduction}
\label{sec:intro}
Over-the-air (OTA) software/firmware update systems are witnessing a huge demand in the automotive market, with the unprecedented transformation towards software-defined vehicles~\cite{e2e-market-mckinsey:2020}. 
%The purpose is to (1) decouple software from underlying hardware for faster (simultaneous) footprint development and wider vendor-ship supply; (2) facilitate the continuous pre- and post-market development of safety and convenience features, like Advanced Driver Assistance Systems (ADAS), \emph{x}-by-wire, infotainment, and Telematics; and (3) importantly, 
An OTA update system is key to reduce the safety risks and maintenance time and cost, when post-market fleet anomalies and vulnerabilities are detected~\cite{software-defined-deloitte:2021,software-defined-aptiv:2020,survey-halder-OTA:2020}. This is of paramount importance for Original Equipment Manufacturers (a.k.a., OEMs), that had to \textit{recall} 35 Million vehicles in 2021 alone, and whose estimated recall losses are around half Trillion dollars by 2024~\cite{upstream-sec-rep:2022}. Nevertheless, the promised benefits of an OTA update system may be doubtful if the system itself is insecure and costly.

Our work is motivated by two main problems. The first is driven by a request from a leader Japanese OEM, whose customers (vehicles' owners) are complaining about the cost of OTA updates, due the extensive bandwidth utilization (using a cellular LTE communication). Indeed, the enormous number of Software Lines of Code (SLoC)---estimated to exceed 100 Millions---in mainstream vehicles~\cite{sw-eat-car:2021,auto-market-mckinsey:2019} can yield up to 1G Byte of daily updates. This makes the OTA system slow and expensive. It is expensive since current OTA update systems hinge into cellular LTE/4G connectivity---while future 5G solutions rates cannot be anticipated~\cite{survey-halder-OTA:2020,fog-OTA-fizza:2019}. It is slow since the vehicle is not always turned on, and thus, not always connected and updating. In search for a convenient solution, we realized another facet to this problem, characterized by the centralization of updates on the OEM's repositories at daily peak hours. The experiments in~\cite{andrade-connected-2017,andrade-scheduling-2019} pointed out that the interference caused by simultaneous vehicle updates in a zone at peak hours can reach the bandwidth limit of a telecommunication cell unit with less than 20 vehicles. In addition, the connection's quality deteriorates significantly for the vehicles and other (mobile) users. Unfortunately, there are few works that tackled this issue at the high-level using fog, edge, or even Blockchain forwarding nodes~\cite{edge-updates-2018,fog-updates-2022,andrade-connected-2017,blockchain-secure-2018,dorri-blockchain-2017}, as we explain in Section~\ref{sec:related}; however, none of them has studied the issue deeply.  

The second motivation is driven by the need for automotive OTA update systems that are secure by design; otherwise, the consequences can be fatal---endangering human lives. In fact, automotive OTA update systems are more complex than computer or mobile update systems for several reasons~\cite{survey-halder-OTA:2020,uptane-karthik:2016}; we mention two main ones here. First, there is a voluminous number of software producers (e.g., Tier 1, Tier 2, and Tier 3 third parties) throughout the supply chain, which makes it cumbersome for the OEM to handle the complex \textit{chain-of-trust}, or retain and maintain all the respective update repositories by itself. Second, current automotive architectures are composed of many layers of software-based \textit{secondary} ECUs, whereas only one ECU, i.e., the \textit{primary} ECU or Telecommunication Unit, has an interface to the cyberspace, e.g., via a cellular connection. This enforces the secondary ECUs to download updates through the primary, which is problematic, from a liability viewpoint, as these ECUs are often provided by different vendors. Despite the plenty of literary works on OTA update security (see Section~\ref{sec:related}), the most solid end-to-end direction we found is \textit{Uptane}~\cite{uptane-karthik:2016,uptane-standard} that uses a \textit{separation-of-roles} scheme for security key signing and verification. The idea is to ensure the chain-of-trust across many OEM's \textit{roles} (tasks), including suppliers and secondary ECUs. Despite the security premise of Uptane, the system model and protocols are not rigorously formalized (publicly).

%These two points require defining a complex chain of trust across the different stakeholders to prevent spoof, sybil, and roll-back attacks (as detailed next).

%Third, heavy-weight cryptographic primitives for authentication and integrity validation are sometimes impractical, and may be avoided, as most ECUs are generally constrained-devices compared to commodity computers and mobiles~\cite{uptane-karthik:2016}.

In this paper, we introduce \acronym: a scalable, efficient, and secure OTA update system for vehicles. \acronym is the first OTA update architecture that proposes using update stations as \textit{Points-of-Presence} for software updates, e.g., integrated with Electric Vehicle (EV) charging stations (\station, see Fig.~\ref{fig:OTA-arch}). This new architecture allows vehicles to reduce the bandwidth utilization and latency of updates up to an order of magnitude, compared with cellular LTE/4G downloads. This boosts the resilience of the system as the update server is no longer single point of attack or failure. In \acronym, the OEM notifies the vehicles with the meta-data of any new update, e.g., via a cellular network. The vehicle can then download the corresponding \textit{cached} update images via an update station, that is operated by a new business entity, possibly different from the OEM or suppliers. Updates in \acronym are much faster than using cellular LTE/4G as vehicles can be wired to the update station through Fast Ethernet, Fiber Optics, or using even a Powerline~\cite{PLC-book-2011} with the EV cables. \acronym also avoids over-utilising the general cellular bandwidth and significantly reduces the overload on repository servers. On the other hand, this architecture opens a huge business model opportunity, and thus establishing automotive software \textit{market place} operators, as in mobile systems, and is only recently noticed in the industry~\cite{OTA-EV-HARMAN:2022}.

In addition to the \acronym architecture, our work is the first published academic work that formalizes an end-to-end OTA update system model and protocol, presents \textit{Liveness} and \textit{Safety} proof sketches (in the Appendix), as well as drives a scalable empirical evaluation. In particular, our experiments, on a real cluster, show that \acronym reduces the update latency and cellular bandwidth utilization to an order of magnitude, as expected theoretically, compared to cellular-based solutions, like \textit{Uptane}.
The experiments also show that \acronym is more resilient to attacks and overloads, as it no longer depends on a single update server.

There are two main challenges in \acronym's architecture: security and storage scalability. The security challenge is referred to using the update station as a new stakeholder in the software value chain. We inspire by Uptane's separation-of-roles scheme to provide and prove a formal chain-of-trust model that covers the end-to-end workflow between the software producer and the secondary ECUs. However, contrary to Update, we formalize the protocol and security abstractions. The storage scalability challenge is caused by the need to support hundreds of vehicle models and brands, which requires a huge storage capacity at the \station. We solve this by only caching the relevant updates of vehicle models that are common in a zone, e.g., based on historical mobility patterns. The intuition is that similar vehicle brands and models are more likely to be common in a country or city, and therefore, subsequent vehicles of the same model will already have the updates cached at the \station. In addition, updates are tagged and bundled at a fine-grained device level, e.g., provided by a supplier universal identifier (SUI). This avoids downloading duplicate updates of the same auto part used in different vehicles and models. This is reasonable since auto parts in most vehicle models are mostly supplied by a handful number of well-known (Tier 1 and Tier 2) suppliers regardless of the brand/model. 

The rest of the paper is organized as follows: 
Section~\ref{sec:case} discusses the case for \acronym.
Section~\ref{sec:related} presents the key related works, and Section~\ref{sec:models} presents the system and threat models. Section~\ref{sec:arch} then presents the architecture and protocol formalization, which is then evaluated in Section~\ref{sec:eval}. Finally, we conclude in Section~\ref{sec:conc}, followed by proofs in the Appendix.

\section{The Case for Using OTA Update Stations}
\label{sec:case}

\acronym uses update stations embedded/collocated with EV stations, which represents a paradigm shift. We discuss the plausibility of this design decision by highlighting the issues current OTA update approaches have, and how using update station can resolve them.

A software-defined vehicle can be seen as complex mobile phone. Software update sizes can range from few KBs to GBs~\cite{andrade2017-one-week,Nissan-update-size,VW-update-size,Ford-update-size,Hyndai-update-size}. Due to the slowness and cost of OTA download paradigms, OEMs tend to rollout updates in large packages on weekly or monthly basis~\cite{Tesla-update-freq,Volvo-update-freq}. We however argue that as OTA techniques get more mature, we would expect smaller update sizes with large frequency, especially when critical safety or security measures are concerned. 
Updates in vehicles are, however, different from mobiles in two main ways of high relevance. The first is that a vehicle's usage pattern is very restrictive: a download can only happen while ignition is turned on, and often requiring \textit{Park} gear mode. The second is that hundreds of software/firmware (for tens of ECUs) are developed by various suppliers, which makes updates' distribution and delivery cumbersome to the OEM: it shifts the OEM from its core innovation business to out of its comfort zone, e.g., cloud storage and secure update delivery business\footnote{We started to see this the industry. E.g, https://www.mapnsoft.com/ now pushes updates directly to cars on behalf of three OEMs.}. We explain how these two aspects affect the cellular and Wi-Fi download paradigms currently used by the industry.

\paragraph{Cellular} performing OTA updates using cellular 3G/4G/LTE technology is slow, costly, and exhausts the bandwidth resources. The slowness is referred to two reasons. First, due to the human mobility patterns, a car may be in use, and thus connect to a cellular network, for one hour daily, which is not sufficient to complete current payloads required by major OEMs~\cite{andrade2017-one-week,Nissan-update-size,VW-update-size,Ford-update-size,Hyndai-update-size}. Worse, cars are mostly active during rush hours when the network is also highly loaded~\cite{cellular-2017connected}. Due to the limited cellular \textit{Physical Resource Blocks} (PRB), shared by mobiles too, cellular networks only scale (with a reasonable interference) to almost 20 car simultaneous zonal downloads. As a consequence of these reasons, a major update ($>1GB$) may take one week to complete~\cite{andrade2017-one-week,andrade-scheduling-2019}. %Despite this, cellular downloads require a dedicated data plan to maintain updates, which is deemed costly by end users as per our industrial experience\footnote{Unfortunately, OEMs do not publish this information for business reasons.}. 
Last but not least, cellular-based downloads exhaust the telecommunication network resources since the number of vehicle-to-server connections is linear to the number of vehicles.

\paragraph{Wi-Fi} Several OEMs realize the above cellular limitations and thus require using Wi-Fi for large updates. Unfortunately, using Wi-Fi for updates is very location-restricted, insecure, and exhausts the network's backhaul resources. In a nutshell, having access to Wi-Fi is not the common case in vehicle mobility: public Wi-Fi car access is not prominent worldwide, while private Wi-Fi often have limited range coverage to underground garages or remote parking lots~\cite{cellular-2017connected}. Importantly, car owners will unlikely keep their cars' ignition on for minutes-to-hours until downloads complete. These two issues also exist in the case where home EV chargers are used. From a security perspective, we, as well as other authors~\cite{andrade-scheduling-2019}, do not recommended to hinge on the awareness and experience, or lack thereof, of average users to maintain their Wi-Fi access points, wildly common to have guessable passwords and open configurations~\cite{password-router:2020}. Finally, as in cellular OTA updates, downloading through Wi-Fi access points exhausts the backhaul network resources, since the number of connections is linear to the number of Wi-Fi access points, which are roughly as numerous as vehicles.

%EV home chargers?  

Our design decision stems from understanding vehicles' mobility pattern and driver convenience. An EV driver can utilize the charging time at an EV station to do the software updates. Since the direct network (fiber, Ethernet, powerline) can practically be an order of magnitude faster than a cellular/Wi-Fi connection, payloads can often get downloaded in a single charging instance (e.g., within few minutes). At the EV station, a driver is likely to be attentive, and it is reasonable to keep the ignition on for such a short period.
%; or the vehicle will be in a passive active state, which permits the update process. 
On the other hand, the number of connections between the download server and user is reduced to a logarithmic order, \emph{i.e.}, proportional to the number of update stations downloading (see Section~\ref{sec:eval}). Interestingly, since update stations are shared update sources now, it is possible to optimise update distribution even further by downloading the common dependencies between different car models only once, while keeping the control with the OEM. For instance, popular telecommunication units, gateways, or intrusion detection systems from major suppliers currently appear in tens of car models from different OEMs. Instead of having encapsulated update bundles by the OEM, tagged and downloaded for each model, it is smart to tag updates by the supplier and reference them in the OEMs bundle to keep control of updates. These tagged updates are downloaded only once by the update station, regardless of the OEM, as long as the supplier version is the same. 

\section{Related Work}
\label{sec:related}

%Future
%The Electric/Electronic (E/E) vehicle architectures~\cite{EE-2004automotive} allowed for a wide range of software (e.g., operating systems, firmware, libraries, and applications) running on top of tens of interconnected computational Electronic control Units (ECUs), via in-vehicle networks like CAN, LIN, MOST, FLexRay, Ethernet, etc~\cite{in-vehicle-networks-2015,networks-nolte2005automotive}. Originally, a software update was typically performed at repair service shops via diagnostic gadgets that are directly connected to the vehicle network via an \textit{On-board Diagnostics} (OBD) port~\cite{diagnostics2014}. Since the repair service was often deemed a trusted entity, and the update mechanism was deprived from any wireless actors, security was not a big concern as long as basic integrity checking is being done~\cite{diagnostics2014,AiroDiag-2012}.

The seminal Over-the-Air (OTA) architectures~\cite{mahmud-secure-2005,microsoft-auto-2008} appeared with the introduction of wireless technologies like WIFI, Bluetooth, cellular LTE, and later DSRC and C-V2X~\cite{off-vehicle-2018networking}. In~\cite{mahmud-secure-2005}, the authors discussed the main security building blocks for a wireless OTA update system like using symmetric and asymmetric keys, SSL, and VPN. Their architecture was very simple, where the OEM sends the vehicle a software download link provided by the supplier. For integrity assurance, they have suggested downloading a software updates twice along with the message digest~\cite{mahmud-secure-2005,double-download-2007}, which was later shown to be redundant and unnecessary~\cite{survey-halder-OTA:2020}. Nevertheless, the basic blocks like using symmetric and asymmetric keys for authentication and message digests inspired most of the subsequent OTA architectures~\cite{survey-halder-OTA:2020,AiroDiag-2012,secup-2016,uptane-karthik:2016, blockchain-secure-2018}, including ours. On the other hand, Tesla (whose code is not published) took advantage of  using of \textit{Virtual Private Networks} (VPN) over WIFI and cellular networks to ensure authentication and confidentiality in their very first design in 2012~\cite{blockchain-secure-2018, zhou2019secure-like-tuf}, in addition to the extensive use of \textit{code signing} to ensure the integrity of updates~\cite{hacking-tesla-OTA-2017}.%\footnote{This information is based on reverse engineering; to the best of our knowledge, Tesla has not published their OTA update technique.}.

%Future
%This functionality has been extended to the Internet, thus exploiting the cloud computing model to enable direct software updates from OEMs to the vehicle in two stages. The first was pushing the updates to the repair offices via the diagnosis gadget that authenticates and validates the integrity of the update before installing it in the vehicle. The second phase excluded this gadget by having the vehicle to immediately connect to the OEM repository and install the software. This has been amde possible thanks tothe cellular LTE connectivity available in modern vehicles. In this vein, many proposals have been suggested for authentication, like using the Public Key Infrastructure, or VPNs. Integrity was also addressed by using RSA or ECC encryption, chained-encryption, or redundancy.

%With the increasing recalls of vehicles due to software failure or vulnerabilities, Over-the-Air software update has become a hot topic and a requirements in the last decade. 
The increasing complexity of the automotive software supply chain have called for secure end-to-end OTA update architectures that considered the off-vehicle (i.e., vehicle to the surrounding) and on-vehicle (mainly ECU to ECU) parts. The community has got inspired by The Update Framework~\cite{TUF-2010} (TUF)---not tailored for automotive---that addresses the different stakeholders in the software update value chain. It considered the entire chain of trust by introducing the concept of \textit{separation of roles}, thus making it easy to verify the authentication and integrity properties at any stage by the relevant stakeholders. TUF however has not tackled the automotive workflow in which the OEM plays a major role in deciding updates, and the fact that updates should target many entities, i.e., ECUs, rather than a single entity. 
Uptane~\cite{uptane-karthik:2016,uptane-standard,uptane-threats-systematic-2022} bridged TUF's gaps by proposing an OEM's \textit{update director}, playing the role of the software update controller of the vehicle. Uptane also extended the TUF's \textit{separation-of-roles} concept to ensure the end-to-end trust of chain including those between ECUs, i.e., \textit{primary} and \textit{secondary} ECUs, inside the vehicle. 
%However, recent studies showed that software updates are vulnerable to other attacks. 
\acronym benefits from both, TUF and Uptane, by extending the \textit{separation-of-roles} concept to cover the suggested network of update stations. In this case, the update director in Uptane still defines the updates to be installed on a vehicle, however, it uses a publish-subscribe scheme to guide the vehicle to securely download the updates through the update stations and the software image inventories. 
%This reduces the bandwidth utilization and download latency significantly, and avoids making the update director a bottleneck or a single point of failure.  

Similar to \acronym, several works have recently addressed Uptane's bottleneck, but from a telecommunication perspective. For instance, the authors in~\cite{edge-updates-2018} and~\cite{fog-OTA-fizza:2019} proposed architectures that predict the bandwidth usage, e.g., using historical (weekly) vehicle patterns, to schedule content sharing between edge/fog devices in the cellular backhaul network, and with the help of vehicle-to-vehicle (V2V) update forwarding. Similarly,~\cite{fog-updates-2022} proposed using fog nodes to push updates to vehicles with the help of \textit{pivot} cars, in a V2V manner. The reason behind using V2V update forwarding is that the edge/fog devices in these works are still centralized in the their respective zone, and lead to interference issues in peak hours, where the vehicles mainly pull updates~\cite{andrade-connected-2017}. In our work, we avoid V2V forwarding since it requires some notion of trust between vehicles---which was not addressed in the aforementioned works, and we argue that the industry is not ready for this approach. Other works like~\cite{blockchain-secure-2018,dorri-blockchain-2017} have proposed, at the high level, using V2V forwarding schemes through blockchains. However, these works have not thoroughly discussed the security aspects as we do in our work, and they have shown that exchanged updates are linear with the number of vehicles.
%Finally, this paper presents the first architecture that proposes using update stations, and thus truly reduces the bandwidth usage and download delays. 

%Future
%In many of the above mentioned works, there are several fine-grained techniques used across the protocols to improve their efficiency and security. For instance, differential updates or update snapshots~\cite{diff-updates-patent, nilsson-secure-2008,uptane-karthik:2016} have been used to enable the progressive incremental of updates, e.g., upon download's failure instead of instantiating the entire update. Hardware-assisted verification~\cite{UniSUF-TEE-2021,HSM-secure-OTA-2011,TPM-2016-evaluation}, e.g., using TPM, HSM, or TEE, have been suggested to harden the verification of used security primitives. A/B double-bank ECUs are used as a recovery method for safe reboot after flashing a software update on an ECU~\cite{uptane-karthik:2016}. Although \acronym can benefit from all these techniques, we do not mention them explicitly in this work being orthogonal to our contributions. 

%Surprisingly, and to the best of our knowledge, this is the first academic work that formalizes the system model and the OTA update protocol, including an empirical evaluation, in a rigorous way.

\section{System and Threat Models}
\label{sec:models}

\subsection{System Model}
\subsubsection{Off-Vehicle model}
A typical software update system model is composed of two main entities: software provider/supplier and end-user device. An automotive software update system model has two main differences. First, the vehicle manufacturer---commonly known as \textit{Original Equipment Manufacturer (OEM)}---employs a \textit{Software Update Director} (SUD) service that is in charge of directing and maintaining the vehicle software updates tailored to different vehicle models---for safety and liability reasons. In this case, (tens of) suppliers provide the software updates of the specific parts directly to the manufacturer or through direct \textit{Image Repositories} (IR). A supplier can yet delegate software image production or maintenance to (a chain of) third-party suppliers. 
The manufacturer's SUD and the supplier's IRs are assumed to be connected via an unreliable network, e.g., the Internet, where packets can be dropped, reordered, or modified. However, we assume that the network eventually delivers packets to their destination. This is feasible since automotive software updates are not required to be real-time.

\subsubsection{In-Vehicle model}
The second main difference is that the end-user device (the vehicle in this case) is composed of several smaller devices known as Electronic Control Units (ECUs), most of which are constrained in computation, memory, communications and features (e.g., security, power, etc.). 
We distinguish between two main classes of ECUs: a \textit{Primary} ECU that has decent capabilities~\cite{continental-Gateway-ECU, nxp-Zone-ECU} and connectivity to the surrounding environment, which allows it to play the role of vehicle update manager; and tens of \textit{Secondary} computationally-constrained ECUs that are connected to the Primary via the in-vehicle network, through which they receive relevant updates. Some vehicle architectures can have more than one Primary or yet another layer of ECUs, e.g., tertiary ECUs, but we exclude these for simplicity. For the same reason, we assume the Primary and Secondaries are connected to the same \textit{Automotive Ethernet} network~\cite{auto-eth-nets:2021} without any gateways. We assume that the in-vehicle network eventually delivers packets to their destination despite transient issues. 
Without loss of generality, we assume that the vehicle has a single Primary ECU that connects it to the Internet.

\subsubsection{Update stations model}
We extend this system model with an \textit{Update Distribution Broker} (UDB): a distributed network of $N$ of vehicle update stations (\textit{UStation}), e.g., associated with \textit{Electric Vehicle} (EV) charging stations. In particular, UStations in the same zone could be connected via \textit{fiber-optic} LAN, to benefit from their high throughput and security; whereas UStations could use a WAN or the Internet. There is no specific network topology that governs the UStations, but we assume a decent level of resiliency. The UDB plays the role of a marketplace for software updates, which can coordinate and deliver updates to vehicles in a fast, efficient, and secure way directly through a wired connection, e.g., while an EV is charging. A wired connection between the UStation and the vehicle has the advantage of high throughput and security. Again, this is convenient if the software cables are bundled with the power cables of the EV charging station or using Power-Line Communication (PLC)~\cite{PLC-book-2011}.

Finally, we assume a business model where OEMs have \textit{service level agreements} (SLAs) with a \textit{software update operator}, e.g., can be the EV station operator, that implements an \textit{Update Distribution Broker} (UDB), and thus, allowing the UStations to host and deliver updates to the fleets.

\subsection{Threat and Adversary Model}
\label{sec:adv-model}
\subsubsection{CIA model}
Our threat model considers the Availability and Integrity properties of the CIA security triad model. The goal is to ensure the delivery of authentic and intact ``packaged" software updates in a reasonable time to their destination vehicles.
Given the large supply-chain discussed in the system model, the main challenge is to scale out the distribution of software updates, by using UStations as edge devices, while maintaining a secure chain-of-trust across the entire system.
Confidentiality is not considered in this paper because its lack does not directly impede our goal above. Nevertheless, when it is essential, e.g., to protect the software intellectual property and reduce malware injection via reverse engineering, one can simply encrypt the payload using off-the-shelf encryption techniques, like RSA or ECC. In the same vein, we do not address software security prior to packaging or during development. 
%, nor with attacks against already installed software outside the scope of the update protocol, e.g., development bugs, buffer flows, exploits, intrusions, etc. 
We also assume highly resilient and secure SUDs and IRs, meaning that these should be impervious to attackers. In the worst case, an attacker may compromise one but not both servers at the same time.

\subsubsection{Threats}
An adversary may attack the vehicle by (1) attempting to install forged malicious updates to control an ECU or the entire vehicle (functionality or performance) through \textit{Man-in-the-Middle} (MITM) or \textit{Spoofing} attacks~\cite{uptane-threats-systematic-2022}; (2) impeding the vehicle update processes by generating compatibility issues across software or ECUs, caused by partial updates (e.g., \textit{Partial-bundle-installation} or \textit{Mix-and-match} attacks~\cite{uptane-threats-systematic-2022}); or (3) preventing a new update by replaying the same update when the vehicle performs the update cycle (i.e., \textit{Replay} attack~\cite{uptane-threats-systematic-2022}). 
%; or (4) subverting the protocol to cause arbitrary behaviour. 
The adversary may attack the availability by dropping, delaying, or corrupting updates, e.g., by modifying the contents (e.g., \textit{Endless-data} and \textit{Mix-and-match} attacks) or timestamps (e.g., \textit{Freeze} or \textit{Rollback} attacks~\cite{uptane-threats-systematic-2022}). A detailed description of possible attacks is described in~\cite{uptane-threats-systematic-2022}.

\subsubsection{Adversary capabilities}
To perform these attacks, we assume a strong adversary that is capable of compromising the system components and network to impede our goal. In particular, the adversary can:
\begin{itemize}
    \item \textit{Compromise the ECUs in the vehicle}. Compromising the primary could compromise the updates on all Secondaries as well if the former is trusted. Compromising the Secondary may often have local effect only, e.g., by preventing updates locally, but it may also prevent other ECU updates to complete if there are dependencies (e.g., that require atomic all-or-nothing updates). 
    \item \textit{Compromise the keys.} This includes compromising the keys of the manufacturer's SUD, Image Repositories, or the UStations in the UDB. We however assume that the adversary cannot compromise the SUD and IR at the same time.
    \item \textit{Intercept the communication channels}. This includes intercepting the channels between all system actors, both off-vehicle and in-vehicle. The former is more likely in wireless communications, e.g., Cellular LTE, 5G, or Wi-Fi. The in-vehicle network is possible through accessing the vehicle via the OBD-II port, USB, or the Telecommunication Unit.
    \item \textit{Compromise the cryptographic keys.} The adversary can compromise the cryptographic keys, e.g., through stealthy attacks or \textit{elevation of privileges}, to perform spoofing attacks or modify the signed hash digests of updates. Nevertheless, it is assumed that the attacker cannot break the used cryptographic primitives, like RSA and ECC keys or hash functions, using brute force.
\end{itemize}

%\begin{figure}[t]
%\centering
%\includegraphics[scale=0.4]{IRS-arch3.png}
%\caption{Intrusion Resilience System (IRS) Architecture.}
%\label{fig:IRS-arch}
%\end{figure}

%\input{uptane.tex}
\section{\acronym Architecture and Protocol}
\label{sec:arch}

\subsection{Abstractions and Symbols}
The message abstractions and symbols used throughout the paper are summarized in the next table for the reader's convenience. %{\color{blue} All manifests contain a \emph{nonce}, not included in the following for readability. All stakeholders store all \emph{nonces} sent and received.}

\begin{table}[h]
\begin{tabular}{p{0.32\columnwidth}p{0.58\columnwidth}}
 \toprule
 Symbol & Meaning\\ [0.5ex] 
 \midrule
 $\theta=(\msf{Meta}, h, e, s, d)$ & update meta-data of software $s$ for ECU $e$, with hash digest $h$, and dependencies list $d$ \\ 
 $\tau_x=(\msf{TS}, t,v)$ & $x$'s timestamp $t$ and version $v$ of  \\ 
 $\mu=(\msf{Manifest}, l,\theta,\tau_\mu)$ & meta-data manifest of an update at location $l$\\
$D=\{\mu_{\delta_1}, ...,\mu_{\delta_m}\}$ & set of related update $\delta$ manifests\\
$\Delta=(\msf{Bundle}, D, \tau_\Delta)$ & manifest of a bundle of updates enclosed in $D$\\
$\gamma=(\msf{Status}, R, \tau_V)$ & Status of installed updates enclosed in $R$\\
$\sigma_E=\{a: a\in E\}$ & set of signature keys signed a message\\
$\sig{x}_{\sigma_E}$ & $x$ \& its hash digest signed by all keys in $E$.\\
 \bottomrule
\end{tabular}
\end{table}

% f and l are almost the same, simplify
% d and x almost the same, simplify

% OLD
% A software update package $\Delta$ is composed of an ordered list of fine-grained differential update \textit{chunks}, i.e., $\delta$, and represented as $\Delta=(\delta_1,\delta_2,...,\delta_m)$.
% Metadata is exchanged between entities during various protocol steps using manifests, represented as $X=(PB,S)$, where $PB$ is the payload block of the manifest, and $S$ is a list of signatures of $PB$.
% $PB$ is represented as $PB=(p,c)$, where $p$ is the actual payload, and $c=(t,v,r)$ is metadata of the payload, composed of a timestamp $t$, a manifest version $v$, and the certificate of the signer $r$.
% The most common type of manifest payload is metadata $p=(f_\Delta,l_\Delta,h_\Delta,v_\Delta,e)$ about an update $\Delta$, where $f$ is the filename, $l$ is the file length, $h$ is the hash, $v$ is the version, and $e$ is the target ECU.
% This payload can be extended to include a list of $\Delta$'s dependencies $d$ and conflicts $cf$, thus becoming $p=(f_\Delta,l_\Delta,h_\Delta,v_\Delta,e,d,cf)$.
% If the payload describes a list of updates, it is represented as $[p]$ as a contraction of $(p_1,p_2,...,p_n)$.
% Other payloads are described in the paper when appropriate.
% The signature list $S=((s_1,sp_1),(s_2,sp_2),...,(s_n,sp_n))$ is a list of pairs $(s,sp)$, where $sp$ is the protocol used to produce $s$.
% For simplification, we represent $S$ as $\langle PB\rangle_{i}$, meaning that entity $i$ created a signature list $S$ by signing $PB$.
% These symbols are summarized on Table~\ref{tab:symbols}.

\begin{figure}[t]
\centering
\includegraphics[width=.7\columnwidth]{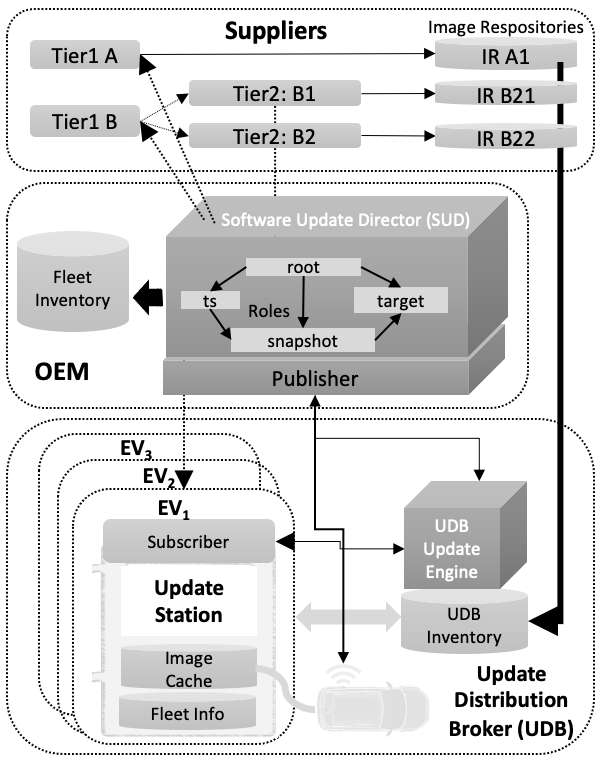}
\caption{The \acronym architecture. 
%A supplier (e.g., Tier 1) produces updates and notifies the OEM. The latter bundles and update metadata with required keys, signatures, and dependencies using the information from the OEM's fleet inventory and time standard server. An EV update station subscribes to the OEM package manager to receive updates corresponding to fleet vehicles in the region. The station downloads the update images via the supplier's image servers after correct delegations. Updates follow a fine-grained tagging scheme to avoid downloading duplicate updates for the same auto part potentially used in several models or brands.
}
\label{fig:OTA-arch}
\end{figure}

\subsection{Architecture overview}
Figure~\ref{fig:OTA-arch} depicts the architecture of \acronym. The architecture is composed of four main parts: the manufacturer's Software Update Director (SUD), the supplier chain of Image Repositories, the EV Station Manager (ESM), and the vehicles. 

%\textbf{Software Update Director (SUD).} The role of the manufacturer SUD is to control the updates of the entire fleet and maintain the updates through the suppliers. For this, the SUD retains an inventory database of the entire fleet information and vehicles specifications, versions, and associated suppliers. The software updates, package dependencies (of different ECUs' software), and the timestamps are all secured through signed manifests with dedicated and redundant cryptographic keys to ensure an end-to-end chain of trust. This is crucial to defend and mitigate more than dozen attacks as mentioned in the threat model, and detailed later in Section~\ref{}. The SUD includes a store for Image Repositories that could retain small software images or simply references to outsourced suppliers.  The SUD communicates with the fleet to push the software updates timely through the Update Update Distribution Broker (UDB) or directly to vehicles through an MQTT publish/subscribe protocol. 
The \textbf{Software Update Director (SUD)} is an entity that is owned by the OEM to control the updates of the entire fleet. For this, the SUD retains an inventory database of the entire fleet information and vehicle specifications, versions, and associated suppliers. The SUD also calculates version dependencies and conflicts to generate safe update lists. The SUD contains roles that represent entities inside the SUD with different responsibilities; when a role writes metadata it always signs it with its own key(s). The roles are: targets---stores the various updates' information; snapshot---creates metadata describing stable software bundles; timestamp---responsible for applying timestamps to targets' and snapshot's metadata, which is used to know if there are new updates; and root---acts as a Certificate Authority (CA) for the other roles.
\textbf{Image Repository (IR).} This is a database where all OEM-associated software is stored, including packages developed in-house and from \textit{N}-Tier suppliers.
The \textbf{Update Distribution Broker (UDB)} is a network of vehicle update stations (UStations), e.g., associated with EV charging stations, and operated by a new entity we call \textit{update operator}. The UDB acts as a CA for its stations, such that the vehicles can attest the owner of each UStation. Each UStation collects identifiers about vehicle models in its corresponding zone, for which updates are downloaded and cached. The UStation pulls from the UE updates corresponding to the identifier, that by its turn subscribes to the corresponding updates provided by the manufacturer's SUD, through the MQTT pub/sub protocol. The UE then redirect the update downloads to the designated UStation. The UE can optimize the cache of the software updates on zonal-basis. It can also predict update usage, i.e., if an update $\Delta$ becomes available for a vehicle type that is frequently seen in its UStations, then it can preemptively publish it.
A \textbf{Vehicle} is composed of a primary and several secondary ECUs, for which software updates are continuously needed.
%The vehicle learns about new updates only by querying the OEM's SUD. The software images can downloaded via the OEM's IR or through the UStations for faster and more efficient bandwidth utilization. 
The \textbf{Supplier} is the main software \textit{update producer}. It stores updates in local or manufacturer-owned Image Repository (IR). As explained later, a direct supplier (called Tier 1) can securely delegate software to other third-party (Tier 2 or Tier 3) suppliers.
Finally, the \textbf{Time Server} is a trusted source of time for all elements considered.

\subsection{Update Initialization and Publishing}
This section presents the \acronym protocol steps for update installation and publishing to the update stations.

\subsubsection{Initialization}
Before dispatching the vehicle to the market, an OEM installs the most recent software versions for all of its parts (i.e., ECUs). The vehicle's model is known by its unique \textit{Vehicle Identification Number (VIN)}, composed of 17 characters. In particular, the very first 11 characters, we call them \textit{Model Identification Number (MIN)}, define the corresponding features, specifications, and manufacturer. %We assume the vehicle's owner has previously obtained a software update service account from the OEM to maintain its vehicle up-to-date.
To keep the vehicle in circulation up-to-date, the OEM retains the updates' meta-data in a \textit{Fleet Inventory} database.
The Fleet Inventory is essential for the OEM to be able to track the circulating fleet and its corresponding software. This is key to pull any new software updates, through software producers, and push them to the vehicles in a timely manner, to reduce recalls or safety incidents.

Therefore, the Fleet Inventory stores the following information: \textit{VIN}, \textit{MIN}, and a list of all software meta-data manifests: $L_e=\{\mu_s=\sig{\msf{Manifest}, l,\theta,\tau_\mu}_\sigma; s\in  S_e \}$, where $S_e$ is the set of software for ECU $e$ corresponding to MIN; $l$ is the $\delta$ location link; $\theta=(\msf{Meta}, h, e, s, d)$ retains the hash digest, ECU, software $s$, and its dependencies, respectively; and $\tau_s=(\msf{TS}, t,v)$ corresponds to the timestamp and version of $s$. All manifests are signed by the corresponding software producer $prod\in \sigma$. Of particular interest, we identify the \textit{last} available update through its TS message $\tau_s^{last}$.\\

\subsubsection{Publishing Updates}
This section presents the protocol used to publish newly generated software updates to the update stations, thus, making the available to be downloaded by the vehicles. \\

\noindent\fbox{%
  \parbox{.95\columnwidth}{%
    \textbf{Step 1.} The software update producer adds a new update $\delta'$ and a \textit{signed} meta-data $\mu'_\delta$ to the Image Repository (IR).
 }
} \newline %fbox

%TODO: maybe define software producer before

Several \textit{software producers} may continuously supply the OEM with updates for the different software installed in the vehicle's ECUs (e.g., in response to a bug, vulnerability, or feature opened in an \textit{Issue Tracking}  platform). 
The producer can be a supplier, third party (e.g., \textit{Tier 2}) or the OEM development team itself. 
When a new software update $\delta'$ is ready to be deployed, the software producer $prod$ places in an Image Repository (IR) both the update $\delta'$ and its corresponding manifest $\mu'_\delta=\sig {\msf{Manifest}, l',\theta',\tau'_\mu}_\sigma$, \textit{signed} by $prod \in \sigma$; where $l'$ represents the update location of $\delta'$ to download from, $\theta'$ collects the integrity data of $\delta'$, and $\tau'_{\mu'}$ retains the current timestamp and version of the meta-data $\mu'_{\delta'}$.
Notice that different IRs---hosted by the producers or the OEM---can retain the same software images; but we only consider one IR in this protocol, for clarity. \\

%\textit{Delegations.} When $\Delta$ is developed by a supplier, a delegation is created; the OEM stores some $\Delta$ that it did not developed, will sign manifests that mention $\Delta$ and its metadata, but the responsible for $\Delta$ and all security around it is the supplier.

\noindent\fbox{ %
  \parbox{.95\columnwidth}{ %
    \textbf{Step 2.} The producer $prod$ sends to the OEM's update director SUD a copy of the meta-data manifest $\mu'_{\delta'}$. The latter validates its authenticity, integrity, and freshness, signs it, and adds it to its Fleet Inventory, if valid. %
 } %
} \newline %fbox

After saving the software update $\delta'$ and meta-data $\mu'_{\delta'}=\sig {\msf{Manifest}, l',\theta',\tau'_\mu}_\sigma$ in IR, the producer sends $\mu'_{\delta'}$ to the OEM's SUD. (The producer may send $\mu'_{\delta'}$ to several OEMs in case their fleet shares the same software, or possibly ECUs.) When the SUD receives $\mu'_\delta$, that subsumes $\theta'=(\msf{Meta}, h', e', s', d')$ and $\tau'_{\delta'}=(\msf{TS}, t',v')$, it validates it by running the following assertions against the last version it has $\tau^{last}=(\msf{TS}, t,v)$.

In detail, the authentication is validated by $\msf{assertAuth}$, Eq.(1), that asserts that $prod$ is one of the signers of the message, i.e.,  $prod \in \sigma$. The freshness is asserted by $\msf{assertFresh}$, Eq.(2), that ensures the time and version of the new update are strictly newer than the \textit{last} version the SUD has in $\tau_s^{last}$. Finally, the integrity is verified by in $\msf{assertIntegrity}$, Eq. (3-4), that ensures the update's download succeeds via link $l$, the hash of the downloaded version matches that of $h'$, and the corresponding ECU $e$ and software $s$ match those in the meta-data manifest.

\begin{align}
    \msf{assertAuth} := &\  prod \in \sigma \\
    \msf{assertFresh} := &\  t' > t^{last} \land v' > v^{last}\\
    \msf{assertIntegrity} := &\ h' =hash(download(l')) \\
                            &\land e = e' \\
                            &\land s=s' 
\end{align}

If these assertions succeeded, SUD signs $\sig{\mu'_{\delta'}}_\sigma$ by appending the signatures of three roles target, timestamp, root to $\sigma=\{\sigma,target, timestamp, root\}$. The target role signature certifies the update data, the timestamp certifies the approval time, and the root certifies the entire roles. These roles can represent different entities or processes at the OEM, which requires separation of responsibilities, and thus require different signatures. Then it adds $\sig{\mu'_{\delta'}}_\sigma$ list of software $L_e$ in the Fleet Inventory. Otherwise, the OEM notifies the producer and asks for the correct update manifests until it succeeds. It is noteworthy that the OEM may or may not perform the necessary quality assurance (e.g., unit testing or validation) in a suitable setting before adding $\mu'_{\delta'}$. We argue that doing this is, however, important for the OEM being the liable entity---contrary to the producer---about any software failures. \\

\noindent\fbox{%
  \parbox{.95\columnwidth}{%
    \textbf{Step 3.} The OEM resolves the possible dependencies of $\delta'$, creates certified update \textit{bundle} $\Delta'$, and then updates its Fleet Inventory with the new \textit{signed} software bundle.
 }
} \newline %fbox

Although the software update $\delta'$ is validated in Step 2, it is still not ready to be pushed to the vehicle because of potential dependencies. 
For this reason, the OEM's SUD has to prepare a bundle $\Delta'$ that contains $\delta'$ and its resolved dependencies, i.e., possibly many updates $\delta_i$, before shipping it to the vehicle. This means that an update $\delta'$ may not be pushed to the vehicle alone. Note that the OEM may need to create a bundle for each vehicle model having different features (e.g., based on its $MIN$).

There are two types of dependencies. The first is a dependency on other ECUs, i.e., if different ECUs have to be updated at once for the new update $\delta'$ on $e$ to function properly. The OEM can resolve these dependencies since---contrary to the software producer---it has visibility over the entire vehicle system, and thus can figure out the possible conflicts or dependencies across ECUs. The second type of dependencies is local, representing the software modules or libraries that the same ECU $e$ must have. These are already listed  by the producer in $d'$, i.e., together with $\theta'=(\msf{Meta}, h', e', s', d')$. The OEM, however, has to make sure the set of dependencies are installed on $e$; otherwise, it bundles them together with $\delta'$ to be shipped to the vehicle. Finally, any dependency $\delta_i$ is assumed to be generated and represented as a typical software update, possibly generated by multiple producers (including the OEM development team), following Step 1 and Step 2 defined above. 

After the list of dependencies ($\delta_1, \delta_2,...,\delta_m$) has been resolved, the SUD creates a bundle $\Delta'=\sig{\msf{Bundle}, D', \tau'_{\Delta'}}_\sigma$. In this bundle, $D'= \{\mu_{\delta'}, \mu_{\delta_1}, \mu_{\delta_2},...,\mu_{\delta_m}\}$ collects all the manifests of the related dependencies of $\delta'$ (included), so that the vehicle can download them as needed. On the other hand, $\tau'_{\Delta'}=(\msf{TS}, t',v')$ assigns the new timestamp and version of the bundle to ensure freshness at the vehicle side (see details later). Finally, the $snapshot$ role's signature is appended to the bundle $\sigma=\{\sigma, snapshot\}$, thus asserting to the vehicle that this bundle is certified by the corresponding OEM.\\

\noindent\fbox{%
  \parbox{.95\columnwidth}{%
    \textbf{Step 4.} The OEM's SUD publishes the manifest $\tau'_{\Delta'}$ of the new bundle $\Delta'$ to all \textit{subscribers}, including the Update Distribution Brokers (UDB) and vehicles, after including and signing the public key of each.
 }
} \newline %fbox

At this stage, the SUD prepares to notify the subscribers about the new update bundle, via the Publish-subscribe service, run by the SUD. The natural subscribers for updates (of a software $s$ corresponding to an ECU $e$) are the OEM's fleet vehicles. In addition, the Update Distribution Broker (UDB) we suggest in this work is a subscriber on behalf of the operated update stations, as explained in Section~\ref{sec:update}. However, the most relevant point to mention here is that the UDB subscribes to receive all the updates corresponding to any vehicle passing by the network of update stations, provided that the vehicle has access to the UDB service. 

Before publishing $\Delta'=\sig{\msf{Bundle}, D', \tau'_{\Delta'}}_\sigma$ to a subscriber $sub$, the UDB's $publish$ role adds the public key of the subscriber and its own public key to $\sigma=\{\sigma, sub, publish\}$. Appending the subscriber's public key and signing it by the OEM's publish role is necessary for the former to certify its eligibility to download the updates listed in $\Delta'$ from the corresponding IRs. Again, having a separate publish role signature is helpful to stop the downloads when needed, i.e., by simply revoking the publish role key (without having to change the other roles if not needed).

Finally, the SUD publishes the bundle update $\Delta'$ to its subscribers. In particular case of the UDB, the notification is sent to the UDB's Update Engine that can later distribute the updates to the relevant update stations internally. On the other hand, the vehicles may also receive the update bundle upon request as described next, in Section~\ref{sec:update}.\\

\noindent\fbox{%
  \parbox{.95\columnwidth}{%
    \textbf{Step 5.} The UDB validates the authenticity, integrity, and freshness of the update bundle $\Delta'$. If this succeeds, it downloads the update images from the respective Image Repositories (IR), and makes them available at the relevant update stations. 
 }
} \newline %fbox

When the UDB's Update Engine receives the update bundle $\Delta'=\sig{\msf{Bundle}, D', \tau'_{\Delta'}}_\sigma$, it first tries to validate its authenticity, integrity, and freshness against the last version it has $\tau^{last}=(\msf{TS}, t,v)$. Authenticity is verified, in Eq. (6), by ensuring the $publish$ and $sub$ keys are included in $\sigma$. Freshness is verified, in Eq. (7), by asserting that the version and timestamps of the received bundle are strictly in the future of the last one UDB has. Integrity is verified by asserting that the hash of $\Delta'$ matches the received signed hash digest.

\begin{align}
    \msf{assertAuth} := &\  \{publish, sub\} \in \sigma \\
    \msf{assertFresh} := &\  t' > t^{last} \land v' > v^{last}
\end{align}

After this, the Update Engine iterates over all $\delta'_i \in D'= \{\mu_{\delta'}, \mu_{\delta_1}, \mu_{\delta_2},...,\mu_{\delta_m}\}$ to validate them, in a similar way to the steps in Eq. (1-5). If validation fails, the Update Engine requests the correct manifests explicitly, until its can validate them. Then it downloads each $\delta_i$ from the corresponding IR mentioned in $\mu_{\delta_m}$, excluding the invalid ones or those that have been downloaded previously. This is possible after authenticating with the IR, by exchanging $\mu_{\delta_m}$ that includes the public key of the UDB $sub$, signed by the OEM. Afterwards, the Update Engine pushes the updates $\delta_i$ and their corresponding manifests $\mu_{\delta_i}$ to the update stations that are subscribed to these updates. The Update Engine can alternatively send the manifests to the corresponding update stations to download the images by themselves.

%Finally, vehicles that opt to download the update images directly from IR (e.g., are not members of the UDB service) can do so in the same manner as the UDB.

\subsection{Update Protocol}
\label{sec:update}
This section presents the \acronym protocol steps followed by the vehicles to download software updates.\\

% In Table~\ref{tab:update_protocol} is presented the protocol the vehicles follow to get updates from the OEM, and described in the following.\\

% \begin{table}[t]
%     \caption{Vehicle update protocol.}
%     \label{tab:update_protocol}
%     \centering
%     \begin{tabular}{c|c}
%     \toprule
%         1) & Connect to EV and authenticate UStation \\
%         2) & Query SUD \\
%         3) & Get updates from UStations \\
%         4) & UStation cache management \\
%     \bottomrule
%     \end{tabular}
% \end{table}

\noindent\fbox{%
  \parbox{.95\columnwidth}{%
    \textbf{Step 6.} The vehicle's primary ECU sends a \textit{report} $R$ describing the vehicle software, by periodically sending $\gamma=\sig{\msf{Status}, R, \tau_\gamma}_\sigma$ to the OEM's SUD. Timestamps and versions have to be signed by \textit{secondary} ECUs if they do not trust the primary.
 }
} \newline %fbox

In this step, the vehicle attempts to check if there are any software updates to download and install. Nevertheless, the vehicle cannot do this with on its own because of the potential dependencies bundled by the OEM. This is common in automotive OTA updates, where the OEM director is in charge of deciding the updates to be installed for each vehicle.

To do this in \acronym, the primary ECU sends a periodic report $R$ about the current status $\gamma=\sig{\msf{Status}, R, \tau_\gamma}_\sigma$ of the software versions installed on the vehicle to the OEM's SUD. The sending frequency can be once per week, day, or even upon every vehicle ignition, but it must be defined for freshness reasons. The primary prepares $\gamma$ as follows: it adds to a list $R$ the timestamp meta-data $\tau_e$ corresponding to all the vehicle ECUs' versions that are currently installed. If $R$ is the same as the previous sent Status message, the primary sends the hash digest of $R$ instead. The primary also prepares $\tau_\gamma=(t_\gamma, v_\gamma)$ by updating its timestamp $t_\gamma$ and using the last Status message's version $v_\gamma$ it has. Finally, the primary signs $\gamma$ by adding its signature $primary$ to $\sigma$ and sends it to the SUD.

In the case where a \textit{secondary} ECU does not trust the primary, which has the only communication interface to the outside world, the secondary has to sign the $\tau_e$ by itself, and \textit{relays} it through the primary while preparing $\sigma$. This has the advantage that the primary cannot lie to the OEM about the last versions installed on the secondary. In addition, the secondary has to send $\tau_e$ at a fixed frequency, e.g., once per week, day, or even upon every vehicle ignition, so that it can raise an \textit{alert flag} to the driver, when the primary delays or discards the OEM's replies of the Status message. \\

\noindent\fbox{%
  \parbox{.95\columnwidth}{%
    \textbf{Step 7.} Upon the receipt of $\gamma=\sig{\msf{Status}, R, \tau_\gamma}_\sigma$ from the vehicle, the SUD validates $\gamma$, identifies the update bundle $\Delta_i$ newer to those reported in $R$, if any, and then encapsulates them in $\gamma'$ as a response to the vehicle.  
 }
} \newline %fbox

When the SUD receives $\gamma=\sig{\msf{Status}, R, \tau_\gamma}_\sigma$ from the primary ECU of vehicle $V$, it first validates its authenticity and hash integrity, which occurs similar manner as before. The exception is freshness validation: if $R$ has been previously seen by the SUD, freshness is ensured if $t^{last}_\gamma < t_\gamma$ and $v^{last}_\gamma \geq v_\gamma$; where $\tau_\gamma=(t_\gamma, v_\gamma)$, and $v^{last}_\gamma$ and $t^{last}_\gamma$ correspond to the last retained Status message's meta-data at the OEM's SUD. The validation of $v_\gamma$ is important since the vehicle should never have a newer version than the OEM. If $\gamma$ is invalid, the SUD will discard the message, and consequently, the response to the $Status$ message will be delayed, and the ECUs at the vehicle will raise an alert to the driver.

Now, the SUD is ready to prepare for the reply $\gamma'=\sig{\msf{Status}, R', \tau_{\gamma'}}_\sigma$. This is done by figuring out all the bundles' meta-data $(\Delta_1, \Delta_2,..,\Delta_k)$ in the Fleet Repository that are in the future of those reported in $R$. If none, i.e., $R$ has been seen before, $R'$ is set to $R$ again. However, this time with a new timestamp $t_\gamma'$ and incremented Status message's version $v_\gamma'=v_\gamma+1$ in $\tau_{\gamma'}=(t_\gamma', v_\gamma')$, so that the ECUs can assert the freshness of \textit{Status} message. (Recall that each update $\delta$ has another $\tau_\delta$ that is used to validate the freshness of the update.) Finally, the SUD's \textit{timestamp} role signature is added to $\sigma$, and $\gamma'$ is sent back to the primary ECU. 

In the case where secondary ECUs do not trust the primary, the SUD has to verify each ECU's timestamp $\tau_e$ in $\gamma$, and then sign each bundle $\Delta$ destined to $e$. This allows the latter to verify the bundle instead of the primary ECU. \\

\noindent\fbox{%
  \parbox{.95\columnwidth}{%
    \textbf{Step 8.} Upon the receipt of $\gamma'=\sig{\msf{Status}, R', \tau_\gamma'}_\sigma$ from the SUD, the vehicle's primary ECU validates $\gamma'$, and becomes \textit{pending} to download the corresponding bundles' images, i.e., when connected to an update station or directly through the IRs.  
 }
} \newline %fbox

When the vehicle's primary ECU receives the Status reply $\gamma'=\sig{\msf{Status}, R', \tau_\gamma'}_\sigma$ from the SUD, it validates its authenticity and hash integrity, as usual. Freshness validation is however done through asserting that $t^{last}_\gamma < t'_\gamma$ and $v^{last}_\gamma \leq v'_\gamma$, where $\tau_\gamma'=(t'_\gamma, v'_\gamma)$, and $v^{last}_\gamma$ and $t^{last}_\gamma$ correspond to the last retained Status message's meta-data at the primary. Note that its okay if the primary did not receive some $\gamma$ versions between $v^{last}_\gamma$ and $v'_\gamma$, since any recent $\gamma$ version will include all the bundles with their dependencies (again other bundles). Thus, it is sufficient for the primary to download the last bundles, in $\gamma$, it does not have, and thus discard any (delayed) $\gamma''$ message whose $t''_\gamma < t^{last}_\gamma$. 

Afterwards, if $\gamma'$ is valid, the primary retains all the meta-data manifest bundles reported in $R'$, in a \textit{pending} state, until it connects to an update station to download them. However, if $v^{last}_\gamma = v'_\gamma$, then $R'$ has been seen before and, therefore, no new updates are available. At this stage, vehicles that do not wish to download the updates through the update stations of the UDB can still download them via typical wireless, e.g., cellular communication.  

In the case where a secondary ECU $e$ does not trust the primary, the latter must send the relevant bundles of $e$, together with $\tau_\gamma'$, to verify them by itself. If this failed, or got delayed to reach $e$, or even the primary has not received a valid $\gamma'$ on time, any of these ECUs can raise an alert to the driver.\\

\noindent\fbox{%
  \parbox{.95\columnwidth}{%
    \textbf{Step 9.} The vehicle connects to an update station, gets authenticated, and downloads each update image $\delta$ included in all the \textit{pending} meta-data bundles.  
 }
} \newline %fbox

When the vehicle stops at an update station, e.g., while charging at an integrated EV station, operated by the update broker UDB, it tries to connect, preferably, via a fast wired interface (e.g., an Ethernet cable that is bundled with the EV electric cable or a \textit{Powerline} cable~\cite{pavlidou2003power}), or via other available communication media, like Wi-Fi.
Then, the primary exchanges its public key, signed by its private key, with the update station to authenticate it---assuming that the vehicle already has a valid account with the UDB operator. If authentication succeeded, the primary sends all its pending meta-data bundles to the update station. Recall that each bundle $\Delta'=\sig{\msf{Bundle}, D', \tau'_{\Delta'}}_\sigma$ may include many meta-data manifests like $\mu^i_{\delta_i}=\sig {\msf{Manifest}, l_{\delta_i},\theta_{\delta_i},\tau_{\delta_i}}_\sigma$ for multiple updates $\delta_i \in D'$, where the update integrity details are in $\theta_{\delta_i}=(\msf{Meta}, h_{\delta_i}, e_{\delta_i}, s_{\delta_i}, d_{\delta_i})$. The update station uses these details to identify the corresponding image $\delta_i$, and thus pushes it to the primary ECU if it is already cached. Otherwise, the update station have to download the update on the spot in the cases of \textit{cache miss} or \textit{unknown} vehicle model, as discussed in the following. Finally, the primary validates the hash digest of $\delta_i$ against the the manifest's digest $h_{\delta_i}$, thus asserting its integrity. \\

\noindent\fbox{%
  \parbox{.95\columnwidth}{%
    \textbf{Step 10.} The primary ECU pushes each update image $\delta_i$ to its corresponding secondary ECU to install them. The secondary may do partial or full verification depending on whether it trusts the primary or not, respectively. 
 }
} \newline %fbox

Prior to pushing updates from the primary to secondary ECUs, two components are required. The first is an in-vehicle communication medium that supports the basic communication and security abstractions used in the protocol above. For simplicity, we consider an \textit{Automotive Ethernet}~\cite{matheus2021automotive} network between the primary and secondaries. Considering other networks is orthogonal to \acronym since we consider the communication channels as a black-box. The second component is a \textit{Flash Bootloader} tool at the secondaries, which is used to handle the received updates and install them. The Flash Bootloader is assumed to have a running \textit{daemon} that has a defined API to be able to communicate with other processes over the network, e.g., with the primary ECU. To avoid pedantic details, we assume that these tools are immutable (i.e., are never updated).

We differentiate between two cases for pushing the updates from the primary to a secondary ECU, considering whether the secondary trusts the primary or not.

\textbf{Trusted Primary.} Assuming the primary is trusted, it can handle the entire software download process as specified in Steps 6-9 above, on behalf of the secondary. The secondary only need to do \textit{partial verification}: verify the key of the primary ECU, i.e., $primary\in \sigma$, and the update image against the received hash $h_{\delta_i}$ in the meta-data $\theta_{\delta_i}=(\msf{Meta}, h_{\delta_i}, e_{\delta_i}, s_{\delta_i}, d_{\delta_i})$ of the manifest $\mu^i_{\delta_i}$. It may also perform basic sanity checks, e.g., error correction or erasure code. If these are valid, the Flash Bootloader daemon installs the update locally and reboots the secondary ECU if needed.

\textbf{Untrusted Primary.} An untrusted primary may lead to integrity or availability issues (and confidentiality if considered). It could violate integrity by forging the software update images or manifests. This would require the secondary to do a \textit{full verification}, which means verifying all the encapsulated messages $\Delta$, $\mu_{\delta}$ $\tau$, $\theta$, and $\sigma$ as the above in Steps 6-9. 
On the other hand, the primary may simply delay or drop the updates destined to the secondary. To address this, the secondary's Flash daemon has to send the \textit{Status} message following a pre-defined frequency, as discussd in Step 6 and 8, and then raise an alert if no valid responses are received on time. 
This ensures that the primary could neither delay or discard the received updates from the OEM beyond the pre-defined period (e.g., defined on safety basis), nor tampering with their content.

%Future: after pushing, there is flashing. If some ECU failed and there are dependencies, there others should roll-back. This needs some atomic broadcast protocol to be done well. What is your take?

%\subsection{Extensions}

\subsection{Caching Mechanism}
\label{sec:caching}
As described in the protocol in Step 9, the vehicle can start downloading the update images once it connects to an update station. However, given the storage capacity of the cache at update stations as well as the mobility patterns of similar vehicle models, an update images may or may not be available at the update station for immediate download. We, thus, consider three interesting cases: \textit{Cache Hit}, \textit{Cache Miss}, and \textit{Unknown Vehicle}, discussed below.

\textit{Cache Hit.} The update station has local caches for the required updates.
This means that the UDB's Update Engine has already pushed the updates to this station, as a result of a previous access that had occurred by a vehicle model requiring the same software. The primary can thus download the software images without any delays, i.e., via a wired connection.

\textit{Cache Miss.} The update station does not have the update images requested by the vehicle. In this case, the vehicle's software are assumed to be subscribed to the UDB's Pub/Sub system, but it lost the images possibly because of the cache scheme. This can happen if any vehicle model that shares the same software had already passed by any update station operated by the UDB. The station can thus pull the updates from the UDB's Update Engine. Although the network connecting the update station is assumed to be fast, e.g., using fiber optics, a cache miss will incur additional delays. Despite this, the download can be one order of magnitude faster than a cellular-based connection, as we show in Section~\ref{sec:eval}.

%Future
%Alternatively, the UE sends an update link to download the update from the corresponding Image Repository. The UE sends a security token each time a download is requested by a UStation.

\textit{Unknown Vehicle.} The current vehicle's model ($MID$) is unknown to the update station. In this case, the station sends a subscribe request to the UDB's Update Engine. If the latter has the $MID$ subscription through another update station, it adds the current station to its list of destinations and sends to it the updates. On the other hand, if the Update Engine does not know about the $MID$ at all (i.e., this model has never been seen at any update station), the UDB subscribes to the OEM's SUD and starts getting updates for that model. Therefore, the protocol is followed to make the updates of this car model supported at the UDB and at this very update station.

\section{Evaluation}
\label{sec:eval}

\subsection{Experimental Setting}
We have implemented \acronym in \textit{Python} and \textit{C}. The core footprint is 1700 LOC, excluding dependencies on external libraries, mainly TUF~\cite{TUF-2010}. 
To evaluate \acronym, we used Emulab~\cite{emulab} that allowed us to use tens of real machines (8-cores with tens of GB of memory) without interference with other applications, while emulating network bandwidth and link delays as needed (via using extra delay machines). We ran our code on Ubuntu 20.04 VM/bare metal machines depending on the experiment.
We set up three types of links with different characteristics for the three typical network use cases suggested in \acronym: cellular from the vehicles to the SUD or IR (5Mbps bandwidth, 30ms latency); cable between UStations and IRs (10Mbps bandwidth, 10ms latency), and direct Ethernet between the UStation and the vehicles (100Mbps bandwidth, 2ms latency).
All measurements include computation times, such as verifying signatures.

\begin{figure}[t]
\includegraphics[width=.95\columnwidth]{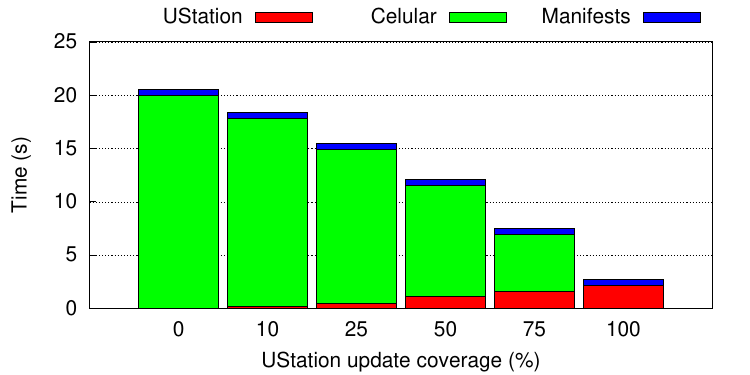}
\caption{Download times considering the percentage of updates served by UStations.}
\label{fig:download_time}
\end{figure}

\subsection{Update download time}

\acronym is attractive to vehicle owners since it uses a direct connection to a UStation, i.e., fast and free, while currently users pay a costly cellular data plan whose speed and bandwidth are capped or limited depending on the region. 
% by the regional signal strength, among other factors that impact wireless connections quality.
In Figure~\ref{fig:download_time} we show this \acronym's advantages for the user by comparing the time taken to download 100MB of updates through cellular link and directly from the UStation.
In this scenario, we consider all update requests to the UStation are hits.
We can see that as we serve a larger percentage of updates from the UStations the time required for downloading the whole bundle decreases significantly, up to $\frac{1}{5}$ of the cellular network time.
We discriminate the time used to fetch the manifests as these are always downloaded via the cellular network.
This figure demonstrates the benefit of using dedicated links for the convenience of the user.

\subsection{Cache management on download time}

\begin{figure}[t]
\includegraphics[width=\columnwidth]{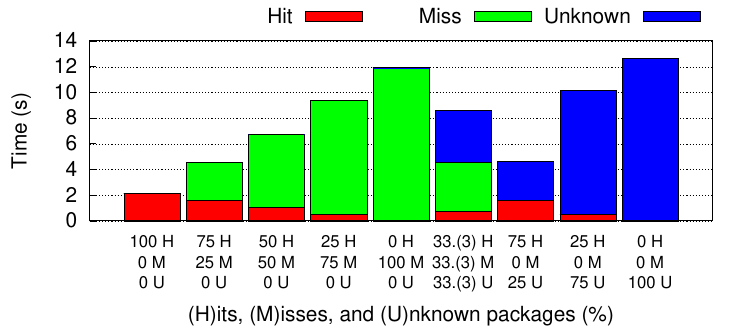}
\caption{Download time comparisons considering various caching events.}
\label{fig:hit-miss}
\end{figure}

Figure~\ref{fig:hit-miss} shows the bandwidth usage of downloading 100MB of updates over the different cache events considered in \acronym: hit, miss, and unknown (see Section~\ref{sec:caching} for details).
Over the X-axis we present several scenarios, where a different percentage of the updates falls under (H)it, (M)iss, and (U)nknown events.
We present a scenario where each event type is dominant, as well as more balanced scenarios.
There are two takeaways from the figure:
1) the update distribution algorithm is paramount for the success of \acronym since we want to prioritize hits for maximum efficiency;
2) the download times between miss and unknown are very similar---this is due to the simplicity of establishing a subscription for a new model, which is the only difference between the processes.
These results reinforce the effectiveness of \acronym, since in the worst case scenario---100\% misses---the download speed is lower than using only the cellular network, while at the same time being free of charge.

\subsection{Resilience of IR to increasing loads and UStations}

\begin{figure}[t]
\includegraphics[width=.95\columnwidth]{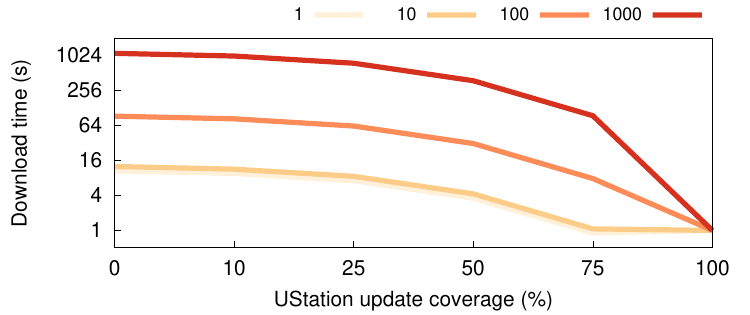}
\caption{The download times based on the number of simultaneous clients.}
\label{fig:numclients}
\end{figure}

In Figure~\ref{fig:numclients}, we show the impact the number of clients have on simultaneously downloading a 100Mb update bundle from the IR.
The X-axis represents the percentage of the bundle that is served by UStations (the remainder is downloaded via the cellular network), while the various lines represent the number of clients simultaneously downloading the bundle.
There is a clear degradation of service as the server gets overloaded with requests when there is little update coverage from the UStations.
Using \acronym means drastically reducing the download times even during peak-usage events, since the load is distributed in the edge instead of centralized in the cloud.
In fact, in a real-world scenario this result is expected to be even worse due to the network congestion and interference from multiple vehicles in close proximity degrading the link quality~\cite{andrade-connected-2017,andrade-scheduling-2019}.

\subsection{Bandwidth Cost}
We estimate the bandwidth cost theoretically since we faced some issues with Emulab while running hundreds of \acronym real instances within hundreds of VMs.
The total update cost is usually paid by owners. It can be modeled as follows: $C_{T} = C_{up} + C_{bwdth}$, where $C_{up}$ is the software update service cost, often offered by the OEM, and $C_{bwdth}= r * (\Delta^a + \mu^a)$ is the estimated bandwidth cost for software $a$ offered by a cellular operator, where $r$ is the telecommunication bandwidth rate. Notice that in state of the art solutions, a software update downloads $\Delta^a + \mu^a$ through the OEM's SUD (that hosts the Image Repository in this case). 

\acronym reduces the $C_{bwdth}$ part, which dominates the update service cost, as the vehicle only downloads $\mu^a$ through the SUD, and $\Delta^a$ via the update station. In his case, it is reasonable to assume the new update service to be equivalent to $C_{up}$, while the bandwidth cost from the update station is assumed to be free. Therefore the total bandwidth cost will be $C_{bwdth} = r * \mu^a$, and thus the relative bandwidth cost will be $\mu^a / (\Delta^a + \mu^a)$.
In our implementation of \acronym, the manifests occupy from a couple to less than a hundred kilobytes, thus confirming that $\Delta^a$ dominates the download cost, and therefore the savings obtained.

\section{Conclusions}
\label{sec:conc}
We have introduced \acronym, a new architecture that makes use of update stations, possibly integrated with EV stations, to reduce the update latency and cellular bandwidth utilization. For this, we have presented a formal OTA update protocol that we have also proved (see Appendix) its integrity, authenticity, and availability properties, defending against known OTA attacks. Our empirical evaluation on a real cluster confirms that the reduction in update latency and cellular bandwidth utilization is one order for magnitude compared with using cellular connectivity alone. Interestingly, \acronym introduces a new business model, by having separate software distribution operators and a market place. In the future, we aim to conduct a dedicated security verification, which is possible now given our formalization. We also aim to use real data traces to understand the mobility patterns over an area near update stations, and thus, better asses the cache misses and hits.

%%%%%%%%%%%%%%%%%%%
%\clearpage
%\newpage
\bibliography{ref.bib}
\bibliographystyle{plain}

%%%%%%%%%%%%%%%%%%%
\appendix

\section*{Proofs}
\label{sec:proofs}

\subsection{Proof methodology}
We provide proof sketches for \acronym by ensuring the Safety and Liveness properties. Safety means \textit{nothing bad happens}, and mainly guarantees the \textbf{Integrity (Int)} and \textbf{Authenticity (Auth)} of the updates in our context. In detail, updates' integrity can be ensured throughout the entire workflow by the following invariants:

\begin{table}[h]
    \centering
    \begin{tabular}{ll}
         S1 (Auth): & an update is only downloaded from trusted producers \\
         S2  (Int): & an update is never modified\\
         S3  (Int): & an update dependency list is never modified\\
         S4  (Int): & an update dependency is never modified\\
         S5  (Int): & a bundle of updates is never modified\\
         S6  (Int): & a primary or secondary ECU never deliver an old update
    \end{tabular}
\end{table}

On the other hand, Liveness means \textit{something good eventually happens}, and mainly guarantees the \textbf{Availability} of a new updates, i.e., a produced update by the producer eventually gets delivered at the corresponding ECU. Availability can be specified throughout reverse walking the entire workflow, ensuring the following invariants:

\begin{table}[h]
    \centering
    \begin{tabular}{ll}
         L1: & a secondary ECU eventually receives new updates via the primary \\
         L2: & a primary ECU eventually receives new updates through the UDB \\
         L3: & the UDB eventually receives new updates through the OEM's SUD \\
         L4: & the SUD eventually receives new updates through the producer\\
         L5: & L1-L4 hold true for any update dependency\\
         L6: & L1-L4 hold true for any update in a bundle
    \end{tabular}
\end{table}

We first recall our trust model. The software producer and OEM are considered trusted entities. The Image Repository IR and UDB are not trusted for safety, but it is reasonable to assume they are trusted for availability due to \textit{Service Level Agreement (SLA)}. Otherwise, the users or OEMs would stop the service. Finally, the primary ECU may be trusted or not, we consider both cases. Finally, any other entity or network is untrusted, although a network eventually delivers its packets (after a certain number of retransmissions).

Finally, we do not explicitly refer to the signatures of OEM's \textit{roles} in the following, as roles are all owned by the same trusted entity OEM---and often used for separation of tasks in the OEM and facilitate revocations.

\subsection{Safety: integrity and authenticity}
We next prove the safety properties. 
\textbf{S1} is ensured as follows: any software producer places a signed version of the software update $\delta$ in an Image Repository (IR) and includes it location $l$ in the manifest $\mu_\delta$ (Step 1). Since $\mu_\delta$ is signed by the producer (Step 1) and then verified by the OEM (Step 2), then UDB and primary ECU (Step 4), and the secondary ECU Step 8 (if the primary is untrusted), this ensures the location $l$ is not tampered with. In addition, the OEM and UDB download the update images through IR (Steps 2 and 5, resp.), using $l$. Then the primary ECU only downloads the image through the UDB or optionally IR (Step 9). Finally, the secondary ECU only receives updates through the primary (Step 10). Therefore, IR is the only source of updates placed by the producer through the link $l$.
\textbf{S2} is ensured as follows: when the producer adds the update image $\delta$ to IR, it adds its signed hash $h_\delta$ to $\theta$ in manifest $\mu_\delta$ (Step 1). The image integrity is then validated against the image at the OEM, UDB, primary, and secondary ECU, if the primary is untrusted, in Steps 2, 5, 9, and 10, respectively. 
\textbf{S3} is ensured as follows: the OEM is the only entity that resolves dependencies of an update $\delta$ set in $d$ as part of $\theta$ and signed (Step 3). Afterwards, the UDB, primary, or secondary ECU, if the primary is untrused, always validate the integrity of the messages received, ensuring that this dependency list is intact (Steps 5, 9, and 10).
\textbf{S4} is ensured as follows: since the OEM only adds the dependencies as updates of the form typical update $\delta$, the dependency itself in never unmodified, follows from S2.
\textbf{S5} is ensured in the same logic as S3, applied to the bundle $\Delta$ instead of $\delta$.

Finally, \textbf{S6} is ensured as follows: to download updates, the primary has to get the manifests for some bundles from the OEM. This happens after the primary sends the $\gamma$ status message that includes the last version $v$ it has about every ECU (Step 6). The OEM then makes sure that the last version $v_l$ it has from that vehicle is greater than $v$ (Step 7). The OEM then increments the $v=v+1$ and replies back to the primary if a new update exist (Step 7). There are two cases where the primary may fool the OEM. The first is sending an old version $v'< v_l$, the OEM will bundle and send all the update manifests up to $v_l$ (Step 7). The primary can thus download all the corresponding updates in the bundle, including some old ones it already has, which leaves no safety impact. The primary can also fool the OEM by sending a new version $v''>v_l$, in which case it is discarded by the OEM since that version must have been originated by OEM already, as per Step 7. Therefore, the primary can only get manifests for new updates form the OEM, for which update images can only be downloaded (Step 9). The secondary ECU receives these update manifest and images through the primary in case it trust it, otherwise, the same validation is followed as in the primary (Steps 8 and 10).

\subsection{Liveness: availability}
We now prove the liveness properties.
\textbf{L1} is ensured as follows: any time the primary has new updates, if its trusted by the secondary, it will push the updates to the latter using the Flash Bootloader daemon, as per Step 10. If the the primary is not trusted, the secondary should have sent its signed last version via the Status message $\gamma$ and set a timer (Step 8). The OEM replies back with it signed manifest message in the bundle $\Delta$ to the primary, who cannot forge it or drop it or delay it for long, since the secondary will raise an alert if the timer is expired (Steps 8 and 10). If the manifest is valid, the secondary also expects to receive the corresponding update images before a timer, otherwise it raises an alert. Note that the timers in this case is set long enough, e.g., can be days, which is arguably reasonable in software updates if there are no critical vulnerability or safety issues.

\textbf{L2} is ensured as follows: the primary sends periodic Status messages, e.g., on each ignition, to the OEM and sets a timer (Step 6). This happens because the vehicle is subscribed to the updates by default. If the OEM did not reply with \textit{valid} bundle manifests in the Status message $\gamma$ before the timer's expiry, the primary raises an alert to the driver. The OEM cannot lie about resolving new updates being the prime trusted entity, and liable, for this (Step 7). However, the primary validates the Status message to ensure that it has not been manipulated by external entities, e.g., a \textit{Man in the Middle} or delayed by the network. Now, since the primary has the manifests, it can download the corresponding update images when it authenticates successfully to a valid UDB update station. The latter should eventually push the updates to primary, from this very station or others in the near future (Step 9). This is reasonable because we assume the UDB operator must guarantee image availability of updates as per the SLA. Again, the primary also validates these new update images across the manifests it has from the OEM. It is noteworthy to recall that the primary can always download the update images from the IR directly, e.g., using its cellular connection.

\textbf{L3} is ensured as follows: the UDB eventually accepts subscriptions from new vehicles stopping at an update station (Step 8), otherwise the SLA is violated. The update station requests the subscription via the UDB Update Engine (internally), and then subscribes to the OEM's SUD, for the same reason. Therefore, the (trusted) OEM will always push new bundled update manifests $\Delta$ to the UDB that validates them in face of external attacks (Step 4) as explained in L2. Now, the UDB can use the link $l$ in the manifest of each $\mu_\delta$ in the received bundle $\Delta$, to download the images from the IR. Since the IR is trusted on availability, as per the SLA contract, the UDB will eventually get the validated new updates against the manifests (Step 5).

\textbf{L4} is ensured as follows: the SUD always receives signed manifests $\mu$ for new updates (Steps 1 and 2), following issues raised in the Issue Tracker Platform, which is managed by an SLA, out of the scope of \acronym. However, the SUD makes sure the received manifests $\mu\delta$ are valid against the images downloaded from IR, via the enclosed download link $l$ in $\mu\delta$ (Step 2). Again, the IR must make the link available as per the SLA service, however, validation is important against external attacks. Finally, the SUD is trusted to bundle the new updates and pushing them to the subscribers (Step 3 and 4), including UDB, and resolve them against old updates in the Status (Step 7) as discussed before. \textbf{L5} is ensured as follows: the (trusted) producer defines the new update dependencies in a signed $\mu_\delta$ (Steps 1 and 2). The OEM validates this manifest and resolves its dependencies (Step 3), which are added to the (signed) bundle $\Delta$. Since these dependencies are represented as updates ($\delta$) as well (Steps 1 and 3), proven in S3 and S4, then they follow the same logic in L1-L4.
Finally, \textbf{L6} is ensured in a similar way to L5, since a bundle $\Delta$ is a mere collection of updates $\delta$ (proven in S5), and therefore, L1-L4 hold for a bundle.

\subsection{Defending against known OTA attacks}

In this section, we map the above security guarantees to the most relevant attacks often discussed in literature~\cite{uptane-karthik:2016,uptane-threats-systematic-2022}. This is only done for the sake of completeness; whereas, a thorough study is a future work. 
%In the following, we demonstrate how the \acronym protocol defends against the most relevant and impactful attacks.

\acronym defends against \textbf{Tampering (T), Spoofing (S), and Arbitrary Software (Ar)} attacks given the manifests' integrity guarantees according to safety proof \textbf{S1}, that describes the full trust-chain from servers, to EV Stations, to vehicle. Since the manifests are signed by the key elements in the chain (T+S+Ar) and contain the hashes of the updates to be installed (T+Ar), an attacker cannot tamper with the update process without being detected (proofs \textbf{S2-6}). If an attacker can obtain control of one of the servers, she may alter at will the data provided by that server (see Section~\ref{sec:adv-model} for details). However, the OTA client will detect that the information coming from both servers (SUD and IR) does not match (T), thus aborting the update.
\textbf{Drop-request and Slow-retrieval} attacks are defended through the ignition-triggered enforced update requests together with availability guarantees \textbf{L1-L6}. Similarly, a \textbf{Partial bundle installation} attack through denial of updates cases an alert to the driver.
%\textbf{Confidentiality}: As described in Section~\ref{sec:adv-model}, we do not address confidentiality in our model. %However, it is possible to setup a dedicated CA for the \acronym stakeholders, distribute the various relevant certificates, and establish TLS connections in every link.
\textbf{Freeze (F), Replay (RE), or Rollback (RO)} attacks are avoided via \acronym due to the liveness properties \textbf{L2}, \textbf{L5}, and \textbf{L6}, where messages will be discarded by the client if these do not present a fresh timestamp (F+RE+RO). Even if the attacker uses messages that have not yet expired, the primary ECU will detect these attacks due to the replayed \emph{nonces} of the signed manifests (F+RE+RO). On the other hand, in \textbf{Mix bundles} attack, if an attacker takes control of one of the server (director or image), she will be able to change the manifest's structure and send to the client and update bundle with conflicting or incompatible packages. However, the client will be able to detect the attack since the two servers' manifest will not match, similar to the integrity guarantees (\textbf{S3-S5}).

Finally, since confidentiality is not our main concern, \textbf{Eavesdropping} attacks are not addressed. Similarly, contrary to~\cite{uptane-karthik:2016,uptane-threats-systematic-2022}, we do not provide any guarantees against \textbf{Endless data} attacks that exhaust the storage/memory resources, or against \textbf{Mix-and-match }attacks that assume keys are compromised.

\end{document}